\documentclass[11pt,a4paper]{article}
\usepackage[utf8]{inputenc}
\usepackage{amsmath}
\usepackage{amsfonts}
\usepackage{amssymb}
\usepackage{amsbsy}

\usepackage{graphicx}
\usepackage[left=3cm,right=3cm,top=2.5cm,bottom=2.5cm]{geometry}
\usepackage{mathrsfs}

\usepackage{tikz}
\usetikzlibrary{shapes.geometric, arrows}

\newcommand{\ex}{e^{x}}

\def\hyph{-\penalty0\hskip0pt\relax}

\title{\textbf{A fast method for pricing American options under the variance gamma model}}
\date{}
\author{Weilong Fu\thanks{Department of IEOR, Columbia University, \texttt{wf2232@columbia.edu}}\ \ \  Ali Hirsa\thanks{Department of IEOR, Columbia University, \texttt{ah2347@columbia.edu}}}

\begin{document}
\maketitle
	
\begin{abstract}
We investigate methods for pricing American options under the variance gamma model. The variance gamma process is a pure\hyph jump process which is constructed by replacing the calendar time by the gamma time in a Brownian motion with drift, which makes it a time\hyph changed Brownian motion. In general, the finite difference method and the simulation method can be used for pricing under this model, but their speed is not satisfactory. So there is a need for fast but accurate approximation methods. In the case of Black\hyph Merton\hyph Scholes model, there are fast approximation methods, but they cannot be utilized for the variance gamma model. We develop a new fast method inspired by the quadratic approximation method, while reducing the error by making use of a machine learning technique on pre-calculated quantities. We compare the performance of our proposed method with those of the existing methods and show that this method is efficient and accurate for practical use.
\end{abstract}

\providecommand{\keywords}[1]{\textbf{\textit{Keywords:}} #1}

\section{Introduction}
Financial models based on L\`evy processes are proposed to overcome the problems of the diffusion models, such as the variance gamma model (VG, \cite{madan_variance_1990}, \cite{madan_variance_1998}), the normal inverse gamma model (NIG, \cite{barndorff1997processes}, \cite{rydberg1997normal}), the tempered stable process (also known as the CGMY model, \cite{carr_fine_2002}), and the variance gamma scaled self\hyph decomposable model (VGSSD, \cite{carr2007self}). They are better at both describing the fat tails of asset returns and matching the implied volatility surfaces in option markets. 

American options are important in the financial markets. There are many markets with American\hyph type options, such as Gold, Silver and options on futures e.g. Crude Oil. They are used in market taking/making, trading, mark\hyph to\hyph model and risk management. However, American options are harder to price because of the early exercise. For the Black\hyph Merton\hyph Scholes (B\hyph M\hyph S) model, \cite{barone1987efficient} proposed a fast approximation of the American options based on the quadratic approximation. Later, \cite{ju1999approximate} elaborated the method of \cite{barone1987efficient} to further reduce its error. However, such a fast approximation method does not exist for the pure jump models. Thus a vast body of literature has discussions on pricing American options under the VG process, its generalization CGMY and even more general L\`evy processes.

A variety of the finite difference methods are based on differential equations. Discretization of the backward partial integro\hyph differential equation (PIDE) with the implicit scheme \cite{hirsa2004pricing} is a standard method for pricing. The Fast-Fourier-Transform (FFT) is used to evaluating the integrals in each time step in \cite{almendral_american_2007}. Some other mutations are \cite{cont_finite_2005} and \cite{wang_robust_2007}. Aside from the backward PIDE are the forward PIDE in \cite{carr2003backward} and the fractional partial differential equation (FPDE) in \cite{cartea2007fractional} and \cite{marom_comparison_2009}, which is specialized in the CGMY model. The finite difference methods are accurate but time-consuming.  
\cite{carr_option_1999} used the FFT to price European options. \cite{lord2008fast} made use of the FFT in multiple time steps to price Bermudan options and further American options.
Monte Carlo simulation can also be used to price American options through Longstaff-Schwartz method \cite{longstaff2001} given the generated samples. In \cite{ribeiro_valuing_2003}, the authors proposed a gamma bridge to speed up pricing American options under VG via simulation. 
Those methods can all be used to price American options, but they are time\hyph consuming. To perform the finite difference method or the FFT, we have to divide the axes of time and the stock price into many small intervals and calculate values on each grid point. To perform simulation, we have to generate a huge amount of sample paths.

We would like to find some new method to improve the speed while keeping the accuracy. One direction is to borrow the idea of the quadratic approximation from \cite{barone1987efficient}.  
In \cite{guo_valuation_2016}, the authors proposed an approximation method based on that idea. In their approach, they first find the exercise boundary of American options through a fixed point system and then solve the approximated equation. However, the approximated equation introduces errors since it cannot completely describe the surface of the premium of the American options. Another way is to learn the option price or some parameters of the price surface as a function w.r.t. all the parameters involved in the model. In \cite{tugce2019}, the authors used deep neural networks to learn the function of the option price w.r.t. the model parameters, but for creating labels for their supervised leaning neural networks, they still need to use a model to create those labels to train their networks.

Our paper is aimed to find a new method for pricing American options under the pure jump model, which improves both speed and accuracy. We will focus on the VG model for simplicity, while it can be generalized to other pure jump models. The method combines the strengths of both quadratic approximation and kernel regression. First, although we start from the PIDE, we avoid dealing with time steps like the finite difference method by the same spirit of the quadratic approximation in \cite{barone1987efficient}. Second, we add a correction term to the approximated equation to reduce the error caused by the approximation step. Third, we employ kernel regression, which is a nonparametric machine learning technique, to estimate the correction term using pre-calculated data. The method does not need as much data as learning the option price surface directly. 



The structure of the paper is as follows: In Section \ref{sec:VG} we do a quick review of the VG model and pricing of European and American options under VG. In Section \ref{sec:simple}, we find a simple way to apply Ju\hyph Zhong method \cite{ju1999approximate} to VG. Even though we did not expect this naive approach to be a solution, we thought it was worth examining it, and our numerical tests show that the error can be somethings within the bid\hyph ask spread but often beyond it. In Section \ref{sec:main}, we elaborate our main approach and summarize the algorithm and give some high\hyph level intuitions. In Section \ref{sec:num} we present the results of numerical experiments and show that the main approach performs well in both speed and error. In Section \ref{sec:conclusion}, we conclude the paper and discuss some possible future research.

\section{The variance gamma model}\label{sec:VG}
Let $b(t;\theta,\sigma)=\theta t+\sigma W(t)$ be a Brownian motion with drift $\theta$ and volatility $\sigma$, where $W(t)$ is a one\hyph dimensional standard Brownian motion. Also, let $\gamma(t;1,\nu)$ be the gamma process with mean rate $1$ and variance rate $\nu$. It has independent gamma increments over intervals of length $h$ with mean $h$ and variance $vh$. 

The three\hyph parameter variance gamma process $X(t;\sigma,\theta,\nu )$ is defined by
\begin{eqnarray*}
X(t;\sigma,\theta,\nu )=b(\gamma(t;1,\nu),\theta,\sigma).
\end{eqnarray*}
The obtained process is a time\hyph changed Brownian motion with drift and its increments have a fat\hyph tailed distribution.

The L\`evy density of the VG process is given by 
\begin{align}
		k(x)=\frac{e^{-\lambda_p x}}{\nu x}1_{x>0}+\frac{e^{-\lambda_n \vert x\vert }}{\nu \vert x\vert}1_{x<0},\label{eq:k}
\end{align}
 where $\lambda_p=\left(\frac{\theta^2}{\sigma^{4}}+\frac{2}{\sigma^2\nu} \right)^{\frac{1}{2}}-\frac{\theta}{\sigma}$ and $\lambda_n=\left(\frac{\theta^2}{\sigma^{4}}+\frac{2}{\sigma^2\nu} \right)^{\frac{1}{2}}+\frac{\theta}{\sigma}$. Also, the characteristic exponent of the VG process is given by
\begin{align*}
	\phi(\xi)=-\frac{1}{\nu}\ln(1+\frac{\sigma^{2}\nu \xi^{2}}{2}-\mathrm{i}\theta \nu \xi )
\end{align*}
such that $\ln \mathbb{E} \left( e^{\mathrm{i}\xi X(t)} \right)=t\phi(\xi)$ holds.

The risk neutral process of the stock price under the variance gamma (VG) model is given by  
\begin{eqnarray}
	S(t)=S(0)\exp((r-q)t+X(t)+{\omega} t),\label{eq:p}
\end{eqnarray}
where $r$ is the risk\hyph free interest rate, $q$ is the dividend rate of the stock, and $\omega=\frac{1}{v}\ln(1-\sigma^2 \nu/2-\theta \nu)$. $\omega$ is calculated such that $\mathbb{ E}(S(t))=S_0\exp((r-q)t)$, which is equivalent with the no\hyph arbitrage condition.

Let $\Theta=\{r,q,T,\sigma,\nu,\theta \}$ be the parameter set. Then the price of a European put option with strike $K$ and maturity $T$ under parameter $\Theta$ is 
\begin{eqnarray*}
	p(S(t),t;K,\Theta )=e^{-r(T-t)}\mathbb{ E}_{t}((K-S(T))^{+}). 
\end{eqnarray*}
According to \cite{madan_variance_1998}, the price of a European put option on a stock given by \eqref{eq:p} is
\begin{eqnarray*}
p(S(0),0;K,\Theta )&=&K\exp(-rT)\Psi\left( -d\sqrt{\frac{1-c_2}{\nu}},-\alpha\sqrt{\frac{\nu}{1-c_2}},\gamma\right)\\
	&&-S(0)\exp(-qT)\Psi\left( -d\sqrt{\frac{1-c_1}{\nu}},-(\alpha+s)\sqrt{\frac{\nu}{1-c_1}},\gamma\right)\\
\end{eqnarray*} 
where
$$
	d=\frac{1}{s}\left( \ln \frac{S(0)}{K}+(r-q)T+\frac{T}{v}\ln\left( \frac{1-c_1}{1-c_2} \right) \right),
$$
$c_1=v(\alpha+s )^2/2$,	$c_2=v\alpha^2/2$, $\alpha=\xi s$, $\xi=\theta/\sigma^2$, and $s=\sigma/\sqrt{1+ \frac{\theta^2 v}{2\sigma^2} }$ and the function $\Psi$ is defined in terms of the modified Bessel function of the second kind and the degenerate hyper\hyph geometric function of two variables (see \cite{madan_variance_1998}).

When the risk neutral dynamics for the stock price is $S(t)$, by its Markov property, the American option is priced by $$P(S(t),t;K,\Theta )=\sup_{t\leq \tau\leq T}\mathbb E_t(e^{-r\tau}(S(\tau)-K)^{+}),$$
where the supremum is taken over all stopping times $\tau$ defined on the probability space with regard to the filtration generated by the stock price $S(t)$. For American put options, at each $t$, there exists a critical stock price $S^{\star}(t)\leq K$, such that
if $S(t)>S^{\star}(t)$, the value of the option is greater than the immediate exercise value and the optimal action is to wait, while if $S(t)\leq S^{\star}(t)$ the value of the option is the same as the immediate exercise value and the optimal action is to exercise the option. In the first quadrant of a two\hyph dimensional space, $\{(S,t):S>S^{\star}(t),0\leq t\leq T\}$ is called the continuation region and $\{(S,t):S\leq S^{\star}(t),0\leq t\leq T\}$ is called the exercise region.

\section{A simple approach for pricing under VG}\label{sec:simple}

We proposed two approaches for approximation. The first is a simple one which makes use of Ju\hyph Zhong method \cite{ju1999approximate}. Ju\hyph Zhong method is used to price American options under B\hyph M\hyph S model. Here we want to test whether the methods for pricing under B\hyph M\hyph S model can be borrowed to the VG model. The steps are: 
\begin{itemize}
\item First, we calculate the difference of American and European options of B\hyph M\hyph S model with the volatility replaced by $\sqrt{\sigma^2(\epsilon )}$ and the dividend replaced by $q-\omega(\epsilon )$ where
	\begin{eqnarray*}
		\sigma^2(\epsilon )&=&\int_{\vert y\vert \leq \epsilon}y^2k(y)dy\\
		\omega(\epsilon )&=&\int_{\vert y\vert \leq \epsilon}(1-e^{y} )k(y)dy
	\end{eqnarray*}
	Here $k(x)$ is the L\`evy density of the VG process. The price of American options is given by Ju-Zhong method.
	\item Then we add the difference to the VG European price to get an approximated VG American price.
	\end{itemize}  
	In Appendix \ref{app:first}, we go over the derivation of $\sigma^2(\epsilon )$ and $\omega(\epsilon )$. We set $\epsilon$ to $0.65$ based on empirical tests\footnote{Thanks to Chengjunyi Zheng, Amir Oskoui, Abhishek Sanghani, and Letian Wang for their effort on this method.}. The approach is very fast thanks to Ju\hyph Zhong method, but our empirical results show it is not ``always'' within the bid\hyph ask spread.

%
%

\section{Development of the main approach}\label{sec:main}
We need a more accurate methodology than the simple one that was introduced in Section \ref{sec:simple}. So in this section, we propose and develop the main idea for pricing American options under VG.
 
From Section \ref{sec:oide} to \ref{sec:correction}, we explain the development of the method from the partial integro\hyph differential equation (PIDE) of VG, including using the quadratic approximation to accelerate calculation and employing nonparametric regression to reduce the error. Section \ref{sec:scalability} introduces a property that simplifies calculation. Section \ref{sec:summary} summarizes the method into an algorithm. Section \ref{sec:insights} gives some insights of the method and explains why it works well.
 
\subsection{From PIDE to OIDE}\label{sec:oide}
It is shown in \cite{hirsa2004pricing} that the price of a European option $p(S,t;K,\Theta )$ and the price of an American option $P(S,t;K,\Theta )$ in the continuation region satisfy this PIDE:
\begin{eqnarray*}
	\int_{-\infty}^{\infty}\left[V(Se^{x},t)-V(S,t)-\frac{\partial V}{\partial S}(S,t)S(e^{x}-1) \right]k(x)dx&&\\
	+\frac{\partial V}{\partial t}(S,t)+(r-q)S\frac{\partial V}{\partial S}(S,t)-rV(S,t)&=&0
\end{eqnarray*}
Here $V(S,t)$ being the price and $k(x)$ is the L\`evy density given by Equation \eqref{eq:k}.

By making changes of the variables, $x=\ln S$, $\tau=T-t$ and $w(x,\tau)=V(S,t)$, we get 
\begin{eqnarray*}
	\frac{\partial w}{\partial x}(x,\tau)&=&S\frac{\partial V}{\partial S}(S,t),\\
	\frac{\partial w}{\partial \tau}(x,\tau)&=&-\frac{\partial V}{\partial t}(S,t),\\
	w(x+y,\tau)&=&V(S e^{y},t),
\end{eqnarray*}
and the following equation
\begin{eqnarray}
	\int_{-\infty}^{\infty}\left[w(x+y,\tau)-w(x,\tau)-\frac{\partial w}{\partial x}(x,\tau)(e^{y}-1)  \right]k(y)dy&&\notag \\
	-\frac{\partial w}{\partial \tau}(x,\tau)+(r-q)\frac{\partial w}{\partial x}(x,\tau)-rw(x,\tau)&=&0. \label{eq:pide_middle}
\end{eqnarray}
Considering $\omega=-\int_{-\infty}^{\infty}(e^{y}-1)k(y)dy$, the equation can be simplified as

\begin{eqnarray}
	\int_{-\infty}^{\infty}\left[w(x+y,\tau)-w(x,\tau) \right]k(y)dy&&\notag \\
	-\frac{\partial w}{\partial \tau}(x,\tau)+(r-q+\omega)\frac{\partial w}{\partial x}(x,\tau)-rw(x,\tau)&=&0. \label{eq:pide}
\end{eqnarray}

The early exercise premium is $${w(x,\tau;K,\Theta)}=P(e^{x},T-\tau;K,\Theta )-p(e^{x},T-\tau;K,\Theta),$$ which is the difference of the price of an American option and a European option, satisfying Equation \eqref{eq:pide} in the continuation region $x>\ln(S^{\star}(T-\tau))$, and equals $K-e^{x}-p(e^{x},T-\tau;K,\Theta)$ in the exercise region $x>\ln( S^{\star}(T-\tau))$.

The finite difference method is accurate but time-consuming because the scheme  makes use of the PIDE and divides the time interval into many steps and has to be solved at each time step. The key idea to accelerate is to get rid of the time axis and just focus on the last step. So we want to approximate the PIDE by an ordinary integro-differential equation (OIDE).

We approximate $w(x,\tau)$ in a similar way as {the quadratic approximation} as shown in \cite{barone1987efficient}. Let $w(x,\tau)=h(\tau)f(x,h(\tau))$, where $h(\tau)=1-e^{-r\tau}$, then
\begin{eqnarray*}
	\int_{-\infty}^{\infty}\left[f(x+y,h(\tau))-f(x,h(\tau)) \right] k(y)dy+(r-q+\omega)\frac{\partial f}{\partial x}(x,h(\tau))&&\\
	-\frac{r}{h(\tau)}f(x,h(\tau))-r{(1-h(\tau))f_{h}(x,h(\tau))}&=&0
\end{eqnarray*}

In practice, $(1-h(\tau))f_{h}(x,h)$ is close to $0$ but not exactly $0$. To solve the equation approximately, we omitted the term $r(1-h(\tau))f_{h}(x,h)$. Meanwhile we added a correction term $\mathcal{E}(x;K,\Theta)$ on the r.h.s. of the equation, meaning that the l.h.s. of the equation is not $0$ exactly. Hence we have
\begin{eqnarray}
	\int_{-\infty}^{\infty}\left[w(x+y,T)-w(x,T) \right] k(y)dy&&\notag \\ 
	+(r-q+\omega)\frac{\partial w}{\partial x}(x,T)-\frac{r}{1-e^{-r T} }w(x,T)&=&\mathcal{E}(x;K,\Theta)\label{eq:oide}
\end{eqnarray}

Let $x^{\star}=\ln(S^{\star}(0))$ be the exercise boundary at maturity. The premium $w(x,T;K,\Theta)$ should satisfies \eqref{eq:oide} on $x>x^{\star}$ (continuation region) and $w(x,T;K,\Theta)=K-e^{x}-p(e^{x},0;K,\Theta)$ on $x\leq x^{\star}$ (exercise region). 

\subsection{Solving the OIDE by parameterization}

In Equation \eqref{eq:oide}, $\mathcal{E}(x;K,\Theta)$ is an correction term. It is close to $0$ compared with the other terms on the l.h.s. In this part, we take it as an arbitrary function that is close to $0$, and seek a way to solve Equation \eqref{eq:oide} for an arbitrary $\mathcal{E}(x;K,\Theta)$. We leave it to section \ref{sec:correction} to determine the value of $\mathcal{E}(x;K,\Theta)$. 
	
Although we get the approximation equation \eqref{eq:oide}, we cannot solve it explicitly due to the integral term. So we consider to solve it numerically. We use an exponential function as an approximation for $w(x,T;K,\Theta)$ in the continuation region, which coincides with the explicit solution of the approximation function in \cite{barone1987efficient}:
\begin{eqnarray}
w(x,T;K,\Theta)=\left\{\begin{array}{cc}
K-\ex-p(\ex,0;K,\Theta ) & x\leq x^{\star}\\
\exp({\lambda} (x-{x^{\star}})+{b} ) & x>x^{\star}\\
\end{array}   \right.\label{func:app}
\end{eqnarray}
where $w(x,T;K,\Theta)$ is set to be continuous at $x={x^{\star}}$. There are three parameters in Equation \eqref{func:app}, but ${b}$ can be calculated from ${b}=\log(K-e^{{x^{\star}}}-p(e^{{x^{\star}}},0;K,\Theta )).$ Thus there are two independent parameters in the approximation function.

After parameterizing the premium $w(x,T;K,\Theta)$, we parameterize the l.h.s. of Equation \eqref{eq:oide}.  Define 
\begin{eqnarray}
	g(x;K,\lambda,x^{\star},\Theta )&=&(r-q+\omega)\frac{\partial w}{\partial x}(x,T;K,\Theta)-\frac{r}{1-e^{-r T}} w(x,T;K,\Theta)\notag\\
	&&+\int_{-\infty}^{\infty}(w(x+y,T;K,\Theta)-w(x,T;K,\Theta))k(y)dy\label{eq:g}
\end{eqnarray}
The parametrized OIDE is $$g(x;K,\lambda,x^{\star},\Theta )=\mathcal{E}(x;K,\Theta).$$
We attempt to make $g(x;K,\lambda,x^{\star},\Theta)$ close to $\mathcal{E}(x;K,\Theta)$ at every $x$ on the region $x>x^{\star}$ by minimizing the loss function w.r.t. $\lambda$ and $x^{\star}$: 
\begin{eqnarray}
	\ell(\lambda,x^{\star};\Theta )=\sum_{i=0}^{{N}}(g({x_i};K,\lambda,x^{\star},\Theta)-\mathcal{E}(x_i;K,\Theta ))^2\label{eq:loss}
\end{eqnarray}

We choose {$N=6$} in our numerical experiments\footnote{based on empirical results}. We also choose {$x_i=x^{\star}+\frac{2i}{N} (\ln(K)-x^{\star})$} which are symmetric w.r.t. $\ln(K)$. The choice is to make Equation \eqref{eq:oide} hold both for in-the-money options and out-of-the-money options. Note that $x^{\star}<K$ always holds for put options so these choices are valid independent of the value of $x^{\star}$.

After we solving the parameters $\lambda$ and $x^{\star}$ that minimize the loss function \eqref{eq:loss}, the approximated price of American put is $$
 P(\ex,0;K,\Theta )\approx \left\{\begin{array}{lc}
K-\ex & x\leq x^{\star}\\
p(\ex,0;K,\Theta )+\exp(\lambda (x-x^{\star})+b ) & x>x^{\star}\\
\end{array}   \right.$$

\subsection{Choosing the correction term $\mathcal{E}(x;K,\Theta )$}\label{sec:correction}
The approximation \eqref{func:app} gives a relation between the premium and the parameters $\lambda$ and $x^{\star}$. Solving the loss function \eqref{eq:loss} gives a relation between the parameters $\lambda$ and $x^{\star}$ and $\mathcal{E}(x_i;K,\Theta )$. If we can determine the relation between $\Theta$ and $\mathcal{E}(x_i;K,\Theta )$, we can link $\Theta$ with the premium.

We can decide the value of $\mathcal{E}(x_i;K,\Theta)$ in the following way. First, we can calculate the price of American options by the finite difference method (in fact any valid current method) and the price of European options by the explicit expression or the FFT, and then obtain the true value of $x^{\star}$ from the finite difference method and $\lambda$ by regressing $\ln(P(\ex,0;K,\Theta )-p(\ex,0;K,\Theta)),~x>x^{\star}$ over $x$ and taking the slope. Then those values of $x^{\star}$ and $\lambda$ make the approximation \eqref{func:app} very close to the true value of the premium. We can consider them optimal parameters. 

Let $x^{\star}(\Theta)$ and $\lambda(\Theta)$ be the functions of optimal parameters depending on the parameter $\Theta$. Then $g(x_i;K,\lambda(\Theta ),x^{\star}(\Theta ),\Theta)$ is the optimal l.h.s. of Equation \eqref{eq:oide}. If we take $\mathcal{E}(x_i;K,\Theta)=g(x_i;K,\lambda(\Theta ),x^{\star}(\Theta ),\Theta)$, it is the optimal r.h.s., and it is obvious that the optimal value of \eqref{eq:loss} is $0$ with the optimal solution $\lambda(\Theta)$ and $x^{\star}(\Theta)$.

Now we know how to choose $\mathcal{E}(x_i;K,\Theta)$, but we have already calculated the prices of American options, and it is meaningless to know $\mathcal{E}(x_i;K,\Theta)$ after the prices to achieve a new pricing method. So we need to employ a flexible machine learning technique to learn the value of $\mathcal{E}(x_i;K,\Theta)$. 

To elaborate, we first calculate the value of $\mathcal{E}(x_i;K,\Theta)=g(x_i;K,\lambda(\Theta ),x^{\star}(\Theta ),\Theta)$ for each $i$ at a group of grid points in the parameter space of $\Theta$. Then we fit the surface of $g(x_i;K,\lambda(\Theta ),x^{\star}(\Theta ),\Theta)$ over $\Theta$ for each $i$ using nonparametric regression. By regression, we assume that $g(x_i;K,\lambda(\Theta ),x^{\star}(\Theta ),\Theta)$ is close to a continuous function w.r.t. $\Theta$. In this way we do not have to calculate $g(x_i;K,\lambda(\Theta ),x^{\star}(\Theta ),\Theta)$ for each $\Theta$ and by doing this we will be speeding up the pricing tremendously.

Call the estimate from regression $\hat{g}_i(K,\Theta)$ for each $i$ and we let $\mathcal{E}(x_i;K,\Theta)=\hat{g}_i(K,\Theta)$ in the loss function \eqref{eq:loss}. By doing so, we use a nearly optimal $\mathcal{E}(x_i;K,\Theta)$ in \eqref{eq:loss} and the solution $\lambda$ and $x^{\star}$ are also close to optimal. 

Moreover, we can use the similar methodology to estimate $\lambda(\Theta)$ and $x^{\star}(\Theta)$ from the pre-calculated quantities and use the estimate as an initial solution in the optimization problem of Equation \eqref{eq:loss} to save time.

\subsection{Scalability of price w.r.t. $S$ and $K$}\label{sec:scalability}

According to the property of American and European options and the definitions of $w(x, 0; K,\Theta )$ and $g(x ;K,\lambda,x^{\star},\Theta )$,
\begin{eqnarray*}
	p(\alpha\,S, 0;\alpha\, K,\Theta )&=&\alpha\, p(S, 0;K,\Theta ),\\
	P(\alpha\,S, 0;\alpha\, K,\Theta )&=&\alpha\, P(S, 0;K,\Theta ),\\
	w(x+\ln \alpha , 0;\alpha\, K,\Theta )&=&\alpha\, w(x, 0; K,\Theta )	.
\end{eqnarray*}

In consequence, the exercise boundary $x^{\star}$ changes along with $\ln(K)$ because $x^{\star}=\inf\{x: P(\ex, 0;K,\Theta )>K-\ex \}$. If we change $K$ to $\alpha\,K$, then ${x^{\star}}$ changes to $x^{\star}+\ln \alpha$ and $g(x+\ln \alpha ;\alpha\,K,\lambda,x^{\star}+\ln \alpha,\Theta )=\alpha\, g(x;K,\lambda,x^{\star},\Theta ).$

$\lambda$ is the slope of $\ln(P(\ex, 0; K,\Theta )-p(\ex, 0; K,\Theta )),x>x^{\star}$ against $x$. It remains unchanged after changing $K$ to $\alpha K$. 

The definition of $x_i$ makes it shift along with $x^{\star}$ and $\ln(K)$. Let $$x_i'=x^{\star}+\ln \alpha+\frac{2i}{N} (\ln(\alpha\,K)-(x^{\star}+\ln \alpha))=x_i+\ln \alpha$$ denote the correspondence  of $x_i$ when we change $K$ to $\alpha K$. Then $$g(x_i' ;\alpha\,K,\lambda,x^{\star}+\ln \alpha,\Theta )=\alpha\, g(x_i;K,\lambda,x^{\star},\Theta )$$

Due to the fact that $\hat{g}_i(K,\Theta)$ is an estimate of $g(x_i;K,\lambda,x^{\star},\Theta )$, we obtain $$\hat{g}_i(\alpha K,\Theta)=\alpha\, \hat{g}_i(K,\Theta)$$

So we do not have to calculate $g(x_i;K,\lambda(\Theta ),x^{\star}(\Theta ),\Theta)$ for different $K$'s. We only need a fixed $K_0$ to estimate $\hat{g}_i(K_0,\Theta)$ and then $$\hat{g}_i(K,\Theta)=\frac{K}{K_0} \hat{g}_i(K_0,\Theta)$$

\subsection{Summary of the main approach}\label{sec:summary}
The first part is pre-calculations, which is done prior to pricing: 
	\begin{itemize}
		\item Choose a group of $\{{\Theta_j}\}_{j=1}^{n}$, where $\Theta=(r,q,T,\sigma,\nu,\theta)$ is the parameter set, calculate prices of American and European options $P(S,0;K_0,{\Theta_j} )$ and $p(S,0;K_0,{\Theta_j} )$ by the finite difference method and the FFT respectively for $1\leq j\leq n$ and $K_0=1000$.
		\item Get the exercise boundary $x^{\star}({\Theta_j} )$ from the finite difference method and regress $\ln(P(\ex,0;K_0,{\Theta_j})-p(\ex,0;K_0,{\Theta_j})),x>x^{\star}$ over $x$ to get the slope $\lambda({\Theta_j} )$ for each ${\Theta_j}$.
		\item Calculate $g({ x_i};K_0,\lambda({\Theta_j} ),x^{\star}({\Theta_j} ),{\Theta_j})$ from Equation \eqref{eq:g} for $1\leq j\leq n$ and $0\leq i\leq N$. 
		\item Store the data.
	\end{itemize}

The second part is the pricing routine: Given the strike $K$, the stock price $S(0)$, and all the parameters $\Theta=(r,q,T,\sigma,\nu,\theta)$:
	\begin{itemize}
		\item Use a nonparametric regression routine to estimate $\hat{g}_i(K,\Theta )$ from $$\frac{K}{K_0} g(x_i;K_0,\lambda(\Theta_j ),x^{\star}(\Theta_j ),\Theta_j),1\leq j\leq N.$$
		\item Minimize the loss function w.r.t. $\lambda$ and $x^{\star}$ $$\ell(\lambda,x^{\star};\Theta )=\sum_{i=0}^{N}(g(x_i;K,\lambda,x^{\star},\Theta)-\hat{g}_i(K,\Theta) )^2$$
		\item Get the price 
\begin{eqnarray*}
	&&P(S(0),0;K,\Theta)\\
	&\approx &\left\{\begin{array}{lc}
K-S(0) & S(0)\leq \exp(x^{\star})\\
p(S(0),0;K,\Theta)+\exp(\lambda (\log(S(0))-x^{\star})+b ) & S(0)>\exp(x^{\star})\\
\end{array}\right.
\end{eqnarray*}
\end{itemize}

\subsection{Insights into the main approach}\label{sec:insights}
\newcommand*\circled[1]{\tikz[baseline=(char.base)]{
            \node[shape=circle,draw,inner sep=1pt] (char) {#1};}}
Figure \ref{fig:frame} shows the framework of the approach. The circled numbers emphasize the most important parts in the method.

First, it transforms the PIDE \eqref{eq:pide} into the OIDE \eqref{eq:oide}, which is step \circled{1} in Figure \ref{fig:frame}. In this step we get rid of the time axis, which costs a lot of time in the finite difference method\footnote{It reduces the calculation time from $O(MN)$ to $O(N)$ where $M$ is the number of time steps}. Meanwhile, we keep a correction term to improve the accuracy.

Second, it parameterizes the OIDE and turns the problem of solving an equation into an optimization problem, which is step \circled{2} in Figure \ref{fig:frame}. To solve an equation on the real line, the unknown object is a function $w(x)$ on the whole real line, which is infinite-dimensional. However, this step provides a mapping from the solution of the premium of American options to the correction term $(\mathcal{E}(x_i;K,\Theta))_{i=0}^{N}$, which is only a $N+1$-dimensional vector. This step is essentially dimension reduction. 

Third, as mentioned earlier the approach employs nonparametric regression to make use of the information from the pre-calculated data, which is step \circled{3} in Figure \ref{fig:frame}. Other approaches consider the problem of solving the price given a set of parameters as a single problem. 
This approach considers solving the solution of the price as a group of problems with different parameter $\Theta$'s. If we make use of the mapping from the price to the correction term $(\mathcal{E}(x_i;K,\Theta))_{i=0}^{N}$ built in this method, we can learn the function $\mathcal{E}(x_i;K,\Theta)$ w.r.t $\Theta$. 

If we summarize the main idea of the method to a high level, it should be that the method reduces the solution of the PIDE \eqref{eq:pide} into a low-dimensional space of the correction term vector, uses a nonparametric machine learning technique to fit the surface in the vector space, and then enhances the estimate to an approximated price curve of American options.

\begin{figure}
	\tikzstyle{arrow} = [thick,->,>=stealth, draw=black]
\tikzstyle{process} = [rectangle, minimum width=2cm, minimum height=0.7cm, text centered, draw=black, text=black]
\tikzstyle{empty} = [rectangle, minimum width=2cm, minimum height=0.7cm, text centered, text=black]
\centering
\begin{tikzpicture}[node distance=1.5cm]
\node (pide) [process] {{PIDE}};
\node (oide) [process, right of=pide, xshift=1.5cm] {{OIDE}};
\node (para) [process, right of=oide, xshift=2.5cm] {{parametrized OIDE}};
\node (loss) [process, right of=para, xshift=2.5cm] {{loss function}};
\node (lam) [process, below of= loss] {{$(\lambda,x^{\star})$}};
\node (olam) [process, below of= para] {{optimal $(\lambda,x^{\star})$}};
\node (app) [process, below of= lam] {approximation};
\node (price) [process, below of= olam] {price};

\node (cor) [process, above of= loss] {{$\mathcal{E}(x;K,\Theta)$}};
\node (ocor) [process, above of=para] {{optimal $\mathcal{E}(x;K,\Theta)$}};

\node (text1) [empty, above of=oide,xshift=-1.5cm,yshift=1.5cm] {\textbf{simplification}};
\node (text2) [empty, above of=ocor] {\textbf{pre-calculation}};
\node (text3) [empty, above of=cor] {\textbf{pricing routine}};

\draw [arrow] (pide) -- node[anchor=north] {\circled{1}} (oide);
\draw [arrow] (oide) -- node[anchor=north] {\circled{2}} (para);
\draw [arrow] (para) -- (loss);
\draw [arrow] (price) -- (olam);
\draw [arrow] (olam) -- (para);
\draw [arrow] (para) -- (ocor);
\draw [arrow] (ocor) --  node[anchor=north] {\circled{3}} (cor);
\draw [arrow] (cor) -- (loss);
\draw [arrow] (loss) -- (lam);
\draw [arrow] (lam) -- (app);

\draw[dashed] (4.2,2.5) -- (4.2,-4);
\draw[dashed] (9,2.5) -- (9,-4);
\end{tikzpicture}
\caption{Framework of our main approach.}
\label{fig:frame}
\end{figure}
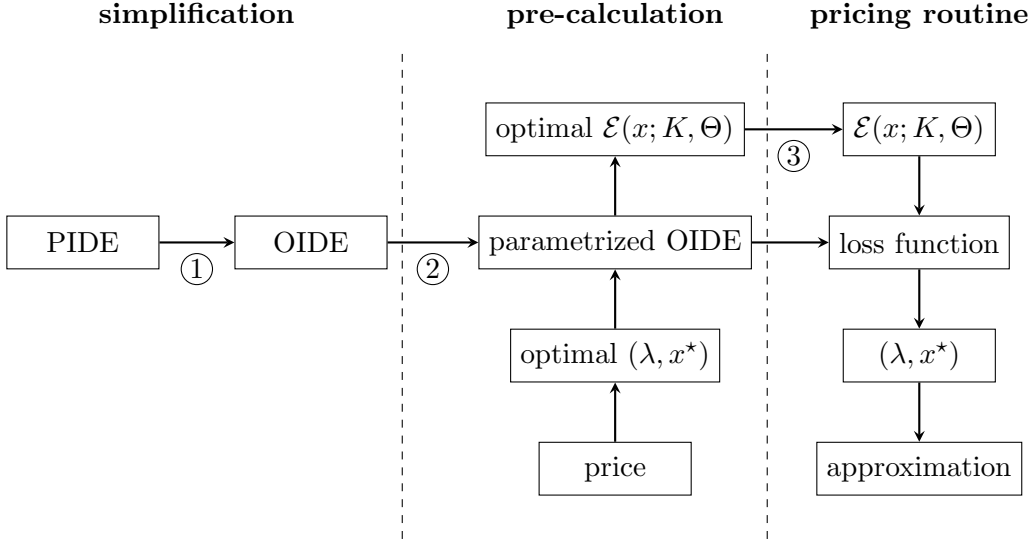

\section{Numerical experiments}\label{sec:num}
 The range of parameters under consideration is
 \begin{eqnarray*}
 	\{\Theta =(r,q,T,\sigma,\nu,\theta):&&0\leq r,q\leq0.1,0.1\leq T\leq 1,\\
 	&&0.1\leq \sigma \leq 0.4,0.1\leq \nu \leq 0.6,-0.5\leq\theta \leq-0.1\}
 \end{eqnarray*}
We pick $S_0=2900$ as it is close to the S\&P 500 Index spot.
	
We compare the following methods in our numerical experiments:
\begin{itemize}
	\item The finite difference method using PIDE in \cite{hirsa2004pricing}. We use the implicit scheme to solve the prices at each time step and Bermudan approach to deal with the early exercise of American options. 
	
	Let $N$ be the number of grid points of $\ln(S)$ and $M$ be grids of time from 0 to $T$. In comparison, we use two versions of finite difference method. On is called FDfine, with $N=3000$ and $M=250$. The other one is called FDcoarse, with $N=800$ and $M=80$. When the grid is finer, the finite difference method is very accurate and can be used as a standard to be compared with. However, that can be time-consuming, so we want to use FDcoarse to test the performance of the finite difference method when we accelerate it with a coarser grid.
	\item MC Simulation with Longstaff-Schwartz method \cite{longstaff2001}. Simulation is a general method of pricing, and Longstaff-Schwartz method is also a general method to deal with the early exercise of American options. The number of time steps is 250 and the number of samples is 1e5.
	\item The simple approach uses Ju-Zhong method (the first proposed approach). 
	\item Proposed main method (the second proposed approach). The grid points where we calculate the optimal parameters $\lambda(\Theta )$ and $x^{\star}(\Theta )$ and then $g(x_i;K,\lambda(\Theta ),x^{\star}(\Theta ),\Theta)$, are the points in the Cartesian product of the following sets:
	\begin{eqnarray*}
		&&r,q\in\{0.01,0.04,0.07,0.1\}\\
		&&T\in\{0.1,0.3,0.5,0.7,0.9,1.1\}\\
		&&\sigma \in\{0.1,0.2,0.3,0.4\}\\
		&&\nu\in\{0.1,0.3,0.5\}\\
		&&\theta\in\{-0.5,-0.3,-0.1\}
	\end{eqnarray*} 

We use kernel regression as the nonparametric method here. The explanatory variable is $\Theta=(r,q,T,\sigma,\nu,\theta )$ and the response variables is $(g(x_i;K,\lambda(\Theta ),x^{\star}(\Theta ),\Theta))_{i=0}^{N}$. The dimensions are $6$ and $N+1$ respectively for the explanatory and response variables. We take $N=6$ in the numerical tests. In Appendix \ref{app:kernel}, we give the details of kernel regression and show how to choose the parameter of the kernel. The ratio of the training set is 75\%. To get a robust choice of the kernel, we repeat the regression for 5 times and take the average of the parameters of the kernels to be the final parameter.

\end{itemize}
The outcomes are shown in Table 1-4 in Appendix \ref{app:results}. All the methods are programmed in C and tested in Matlab on an Intel i7-6820HQ, 2.70GHz. As we can see, the main approach achieves a good balance between small error and fast speed among the methods. The first method (JZ) is usually the fastest, but our main approach has a much smaller error. Also, our main approach is much faster than the finite difference method and the simulation method. Even if we accelerate the finite difference to about 10 times slower than the main approach (FDcoarser), it still has a slightly larger error when $T\leq 0.5$. When $T=1$, the main approach doesn't performs as good as when $T\leq 0.5$, the reason is that when $T$ is larger, the true curve of the premium of American options can not be approximated by the exponential function as well as when $T$ is smaller. 

\section{Conclusion and future work}\label{sec:conclusion}
In this paper we proposed a fast and practical method for pricing American options under the VG model. This method can be viewed from two sides. On one side, it solves an approximated equation with a correction term estimated from the pre-calculated data. On the other side, the optimization routine provides a mapping from the surface of the premium to the vector of the correction terms, which lies in Euclidean space and is easy to estimate. The mapping converts a pricing problem to an easy machine learning problem. 

For future work, option prices in many financial models involving diffusion and jumps can be described with a PDE or PIDE. When we want a fast approximation method for that model, the same idea of the main approach can be applied to NIG, CGMY, and VGSSD. Also, this method is a numerical pricing method. A highly-accurate closed-form approximation solution of the PIDE of the VG model is still attracting.

\medskip
 
\bibliographystyle{abbrv}
\bibliography{references.bib}

\newpage
\appendix
\section*{Appendices}
\section{Development of the first proposed method}\label{app:first}
This part follows \cite{hirsa2016computational}. We can split the integral term in \eqref{eq:pide_middle} into two terms, the integrals on $\vert y\vert \leq \epsilon$ and $\vert y\vert > \epsilon$ respectively.

In the region $\vert y\vert \leq \epsilon$, 
$$w(x+y,\tau)=w(x,\tau)+y\frac{\partial w}{\partial x}(x,\tau)+\frac{y^2}{2} \frac{\partial^{2} w}{\partial x^{2}}(x,\tau)+O(y^{3})$$
and $$e^{y}=1+y+\frac{y^2}{2}+O(y^{3}).$$
Using those two approximations, we get  

\begin{eqnarray*}
	&& \int_{\vert y\vert \leq \epsilon}\left[w(x+y,\tau)-w(x,\tau)-\frac{\partial w}{\partial x}(x,\tau)(e^{y}-1)  \right]k(y)dy\\
	&=&\int_{\vert y\vert \leq \epsilon}\left[\frac{y^2}{2}\frac{\partial^{2} w}{\partial x^{2}}(x,\tau)-\frac{y^2}{2}\frac{\partial w}{\partial x}(x,\tau)+O(y^{3}) \right]k(y)dy\\
	&\approx & \int_{\vert y\vert \leq \epsilon}\left[\frac{y^2}{2}\frac{\partial^{2} w}{\partial x^{2}}(x,\tau)-\frac{y^2}{2}\frac{\partial w}{\partial x}(x,\tau) \right]k(y)dy\\
\end{eqnarray*}
Define $\sigma^{2}(\epsilon)=\int_{\vert y\vert \leq \epsilon}y^{2}k(y)dy$ and we get $$\int_{-\infty}^{\infty}\left[w(x+y,\tau)-w(x,\tau)-\frac{\partial w}{\partial x}(x,\tau)(e^{y}-1)  \right]k(y)dy \approx \frac{1}{2} \sigma^{2}(\epsilon)\left( \frac{\partial^{2} w}{\partial x^{2}}(x,\tau)-\frac{\partial w}{\partial x}(x,\tau)\right) $$

In the region $\vert y\vert > \epsilon$, 
\begin{eqnarray*}
	&& \int_{\vert y\vert > \epsilon}\left[w(x+y,\tau)-w(x,\tau)-\frac{\partial w}{\partial x}(x,\tau)(e^{y}-1)  \right]k(y)dy\\
	&=&\int_{\vert y\vert > \epsilon}\left[w(x+y,\tau)-w(x,\tau)  \right]k(y)dy+\frac{\partial w}{\partial x}(x,\tau)\omega(\epsilon ) \\
\end{eqnarray*}
where $w(\epsilon )=\int_{\vert y\vert >\epsilon}(1-e^{y})k(y)dy$.

Combine the two parts of integrals and put them back to Equation \eqref{eq:pide_middle}, and we get
\begin{eqnarray}
	\frac{1}{2}\sigma^{2}(\epsilon)\frac{\partial^{2} w}{\partial x^{2}}(x,\tau) +\int_{\vert y\vert >\epsilon }\left[w(x+y,\tau)-w(x,\tau) \right]k(y)dy&&\notag \\
	-\frac{\partial w}{\partial \tau}(x,\tau)+(r-q+\omega(\epsilon)-\frac{1}{2}\sigma^{2}(\epsilon) )\frac{\partial w}{\partial x}(x,\tau)-rw(x,\tau)&=&0.\label{eq:pide_ap}
\end{eqnarray}
If we omit the integral term in Equation \eqref{eq:pide_ap}, we can get a B\hyph M\hyph S equation
\begin{eqnarray*}
		-\frac{\partial w}{\partial \tau}(x,\tau)+\frac{1}{2}\sigma^{2}(\epsilon)\frac{\partial^{2} w}{\partial x^{2}}(x,\tau) +(r-q+\omega(\epsilon)-\frac{1}{2}\sigma^{2}(\epsilon) )\frac{\partial w}{\partial x}(x,\tau)-rw(x,\tau)&=&0.
\end{eqnarray*}
It describes the option price of a stock with volatility $\sqrt{\sigma^2(\epsilon )}$ and dividend $q-\omega(\epsilon )$. So we decide to use the premium of this B\hyph M\hyph S model to approximate the premium in the VG model.

\section{Kernel regression}\label{app:kernel}
Kernel regression is a nonparametric machine learning technique that is used to find a non-linear relationship between a pair of variables $x$ and $y$. Both $x$ and $y$ can be vectors. Let $d_{x}$ and $d_{y}$ be the dimensions of $x$ and $y$. Suppose we collect data $x_1,x_2,\dots,x_n$ and $y_1,y_2,\dots,y_n$ and want to find a suitable estimate of $y$ given $x$.

First, to perform kernel regression, we need a kernel function $\kappa (x',x'')$, where $x'$ and $x''$ are two points in the space of $x$. Then the estimate $\hat y=f(x)$ given $x$ is 
\begin{align}
	f(x)=\frac{\sum_{i=1}^{n}\kappa(x,x_i)y_i}{\sum_{i=1}^{n}\kappa(x,x_i)}.\label{eq:reg}
\end{align}

Second we need to choose a suitable kernel function $\kappa(x',x'')$ to get a good estimation. The Gaussian kernel is usually a good choice, i.e., $$\kappa_{a}(x',x'')=\exp(-\sum_{j=1}^{d_{x}}a_j (x'_{j}-x''_{j})^{2}),$$ where $a_j,1\leq j\leq d_{x}$ are positive numbers and $x'_j$ and $x''_j$ are the $j$th component of the vectors $x'$ and $x''$. 

There are different ways to measure the performance of fitting. One way is to define a loss function and choose parameters by optimization. For example, if the components of $y$ are similar, a reasonable loss function can be defined as $$\ell(a)=\sum_{i=1}^{n}\Vert y_i-\hat y^{S}_i(a)\Vert^{2} ,$$ where $\Vert \cdot\Vert $ is the Euclidean norm and $$\hat y^{S}_i(a)=\frac{\sum_{i\in S}\kappa_{a}(x,x_i)y_i}{\sum_{i\in S}\kappa_{a}(x,x_i)}$$ is the estimate of $y$ given $x_i$ and is also a function of $a$. $S$ is a subset of $\{1,2,\dots,n\}$ chosen randomly and is served as the training set. This step aims to avoid overfitting. By minimizing $\ell(a)$, we can get a suitable kernel function $\kappa_a(x',x'')$ for prediction using Equation \eqref{eq:reg}. 

Finally, we can repeat the second step for several times due to the randomness of $S$ and take the average of $a$ for robustness.

\section{Results of the numerical experiments}\label{app:results}

\begin{table}[p]
	\begin{tabular}{|ccc|ccccc|}
	\hline
	r & q & K & FDfine & FDcoarse & main & simulation & simple\\
	\hline 
0.10 & 0.01 & 2600 & 141.939 & 141.594 & 141.801 & 139.954 & 135.297 \\ 
0.10 & 0.01 & 2800 & 198.588 & 198.301 & 198.886 & 195.562 & 192.670 \\ 
0.10 & 0.01 & 3000 & 272.532 & 272.391 & 273.172 & 265.878 & 269.961 \\ 
0.10 & 0.01 & 3200 & 368.504 & 368.685 & 368.549 & 361.026 & 372.550 \\ 
0.05 & 0.05 & 2600 & 156.314 & 156.145 & 156.433 & 155.903 & 156.212 \\ 
0.05 & 0.05 & 2800 & 217.980 & 217.979 & 218.195 & 217.055 & 218.704 \\ 
0.05 & 0.05 & 3000 & 297.861 & 298.157 & 298.187 & 295.774 & 300.286 \\ 
0.05 & 0.05 & 3200 & 400.214 & 401.009 & 400.580 & 399.892 & 405.526 \\ 
0.01 & 0.10 & 2600 & 184.019 & 183.824 & 184.156 & 186.316 & 184.135 \\ 
0.01 & 0.10 & 2800 & 256.889 & 256.903 & 256.947 & 257.258 & 256.916 \\ 
0.01 & 0.10 & 3000 & 351.540 & 351.923 & 351.443 & 351.943 & 351.397 \\ 
0.01 & 0.10 & 3200 & 473.366 & 474.375 & 472.958 & 472.758 & 472.931 \\ 
\hline
& & RMSE & - & {0.429} & {0.291} & {3.224} & {3.378} \\ 
& & MAE & - & 1.010 & 0.640 & 7.479 & 6.642 \\ 
& & CPU(s) & 5.270 & 0.129 & 0.009 & 5.479 & 0.004 \\
\hline
	\end{tabular}
	\caption{Values of American puts. $S_0=2900$, $T=0.5$, $\sigma=0.1$, $\nu=0.6$ and $\theta=-0.5$. RMSE is the root of mean squared errors. MAE is the maximum absolute error. CPU is the mean computing time. }
\end{table}

\begin{table}[p]
	\begin{tabular}{|cccc|ccccc|}
	\hline
$\sigma$& $\nu$ & $\theta$ & K & FDfine & FDcoarse & main & simulation & simple\\
	\hline 
0.10 & 0.10 & -0.50 & 2800 & 26.961 & 27.062 & 27.040 & 26.882 & 26.479 \\ 
0.40 & 0.60 & -0.50 & 2800 & 80.429 & 80.563 & 80.484 & 79.724 & 81.241 \\ 
0.10 & 0.60 & -0.10 & 2800 & 12.088 & 12.114 & 12.596 & 11.976 & 11.927 \\ 
0.40 & 0.10 & -0.10 & 2800 & 71.324 & 71.735 & 71.317 & 72.177 & 71.451 \\ 
0.10 & 0.10 & -0.50 & 2900 & 53.119 & 53.372 & 53.191 & 52.564 & 52.248 \\ 
0.40 & 0.60 & -0.50 & 2900 & 100.103 & 100.741 & 100.121 & 98.937 & 101.765 \\ 
0.10 & 0.60 & -0.10 & 2900 & 24.642 & 23.303 & 24.863 & 24.880 & 24.680 \\ 
0.40 & 0.10 & -0.10 & 2900 & 110.992 & 111.634 & 110.895 & 110.727 & 111.251 \\ 
0.10 & 0.10 & -0.50 & 3000 & 104.401 & 104.780 & 104.275 & 103.621 & 102.507 \\ 
0.40 & 0.60 & -0.50 & 3000 & 130.478 & 132.460 & 130.763 & 129.990 & 132.738 \\ 
0.10 & 0.60 & -0.10 & 3000 & 99.999 & 99.990 & 100.000 & 100.000 & 96.103 \\ 
0.40 & 0.10 & -0.10 & 3000 & 170.431 & 170.888 & 170.459 & 170.141 & 171.060 \\ 
\hline
& & & RMSE & - & {0.772} & {0.189} & {0.575} & {1.549} \\ 
& & & MAE & - & 1.982 & 0.508 & 1.165 & 3.896 \\ 
& & & CPU(s) & 5.085 & 0.129 & 0.010 & 5.340 & 0.017 \\
\hline
	\end{tabular}
	\caption{Values of American puts. $S_0=2900$, $T=1/12$, $r=0.05$ and $q=0.01$. RMSE is the root of mean squared errors. MAE is the maximum absolute error. CPU is the mean computing time. }
\end{table}

\begin{table}[p]
	\begin{tabular}{|cccc|ccccc|}
	\hline
$\sigma$& $\nu$ & $\theta$ & K & FDfine & FDcoarse & main & simulation & simple\\
	\hline 
0.10 & 0.10 & -0.50 & 2800 & 59.653 & 59.862 & 59.513 & 59.787 & 57.920 \\ 
0.40 & 0.60 & -0.50 & 2800 & 185.871 & 186.004 & 186.024 & 187.613 & 188.529 \\ 
0.10 & 0.60 & -0.10 & 2800 & 30.065 & 30.105 & 30.303 & 29.745 & 29.188 \\ 
0.40 & 0.10 & -0.10 & 2800 & 158.556 & 159.291 & 158.513 & 157.752 & 158.927 \\ 
0.10 & 0.10 & -0.50 & 2900 & 94.364 & 94.800 & 94.134 & 93.273 & 91.772 \\ 
0.40 & 0.60 & -0.50 & 2900 & 218.131 & 218.341 & 218.325 & 219.145 & 222.496 \\ 
0.10 & 0.60 & -0.10 & 2900 & 52.210 & 52.428 & 52.392 & 51.713 & 51.398 \\ 
0.40 & 0.10 & -0.10 & 2900 & 205.451 & 206.251 & 205.350 & 203.691 & 206.040 \\ 
0.10 & 0.10 & -0.50 & 3000 & 143.833 & 144.524 & 143.769 & 142.954 & 140.162 \\ 
0.40 & 0.60 & -0.50 & 3000 & 255.864 & 256.189 & 256.130 & 255.742 & 262.532 \\ 
0.10 & 0.60 & -0.10 & 3000 & 99.999 & 100.454 & 100.000 & 100.000 & 95.936 \\ 
0.40 & 0.10 & -0.10 & 3000 & 260.438 & 261.227 & 260.350 & 260.751 & 261.350 \\ 
\hline
& & & RMSE & - & {0.495} & {0.163} & {0.924} & {3.069} \\ 
& & & MAE & - & 0.800 & 0.265 & 1.760 & 6.668 \\ 
& & & CPU(s) & 5.161 & 0.126 & 0.009 & 5.567 & 0.007 \\ 
\hline
	\end{tabular}
	\caption{Values of American puts. $S_0=2900$, $T=1/4$, $r=0.05$ and $q=0.01$. RMSE is the root of mean squared errors. MAE is the maximum absolute error. CPU is the mean computing time. }
\end{table}

\begin{table}[p]
	\begin{tabular}{|cccc|ccccc|}
	\hline
$\sigma$& $\nu$ & $\theta$ & K & FDfine & FDcoarse & main & simulation & simple\\
	\hline 
0.10 & 0.10 & -0.50 & 2700 & 99.153 & 99.283 & 98.397 & 98.443 & 95.105 \\ 
0.40 & 0.60 & -0.50 & 2700 & 388.210 & 388.464 & 388.304 & 386.251 & 393.837 \\ 
0.10 & 0.60 & -0.10 & 2700 & 51.101 & 51.065 & 50.808 & 50.683 & 47.227 \\ 
0.40 & 0.10 & -0.10 & 2700 & 299.414 & 300.591 & 299.140 & 299.877 & 299.372 \\ 
0.10 & 0.10 & -0.50 & 2900 & 173.557 & 174.300 & 171.605 & 171.854 & 166.993 \\ 
0.40 & 0.60 & -0.50 & 2900 & 477.476 & 478.033 & 477.634 & 476.004 & 489.716 \\ 
0.10 & 0.60 & -0.10 & 2900 & 106.615 & 106.757 & 106.492 & 105.221 & 101.473 \\ 
0.40 & 0.10 & -0.10 & 2900 & 398.183 & 399.491 & 397.314 & 395.952 & 398.154 \\ 
0.10 & 0.10 & -0.50 & 3100 & 279.825 & 281.131 & 277.009 & 277.950 & 270.715 \\ 
0.40 & 0.60 & -0.50 & 3100 & 578.029 & 578.940 & 578.297 & 580.304 & 599.734 \\ 
0.10 & 0.60 & -0.10 & 3100 & 208.763 & 209.230 & 208.261 & 205.385 & 203.510 \\ 
0.40 & 0.10 & -0.10 & 3100 & 511.520 & 512.885 & 509.924 & 511.852 & 511.607 \\ 
\hline
& & & RMSE & - & {0.850} & {1.160} & {1.754} & {8.486} \\ 
& & & MAE & - & 1.365 & 2.816 & 3.378 & 21.705 \\ 
& & & CPU(s) & 5.130 & 0.125 & 0.010 & 5.758 & 0.004 \\
\hline
	\end{tabular}
	\caption{Values of American puts. $S_0=2900$, $T=1$, $r=0.05$ and $q=0.01$. RMSE is the root of mean squared errors. MAE is the maximum absolute error. CPU is the mean computing time. }
\end{table}

\end{document}